\documentstyle[12pt]{article}
\begin{document}
\title{QUANTUM SUPERSTRINGS AND QUANTIZED FRACTAL SPACE TIME}
\author{B.G. Sidharth\\
B.M. Birla Science Centre, Adarshnagar, Hyderabad - 500 063, India}
\date{}
\maketitle
\begin{abstract}
Though Quantum SuperString Theory has shown promise, there are some puzzling
features like the extra dimensions, which are curled up in the Kaluza-Klein
sense. On the other hand a recent formulation of what may be called Quantized
Fractal Space Time leads to a surprising interface with QSS - we deal with
the Planck length, the same number of extra dimensions and an identical
non-commutative geometry. It is shown that this is not accidental. On the
contrary it gives us insight into the otherwise inexplicable features of
QSS.
\end{abstract}
\section{Introduction}
At the outset it must be pointed out that Quantum Mechanics (QM) and
Quantum Field Theory (QFT) operate in the
Minkowski space time - this is a differentiable manifold and space time
points, infact point particles are legitimate. However this contradicts the
very spirit of QM - arbitrarily small space time intervals imply arbitrarily
large momenta and energies. Yet both QM and QFT have co-existed with this
unhappy situation\cite{r1}. From this point of view, two approaches are
more satisfactory. One is string theory and the other is the discrete space
time formulation. The former has evolved into the present theory of Quantum
SuperStrings (QSS) while the latter leads to a formulation in terms of
stochastic, fractal space-time. These two approaches appear very different,
but as we will now show, they share several strikingly similar characteristics.
This will throw light on some features of string theory which appear very
puzzling, like the extra dimensions and the non-commutative geometry aspect.
\section{String Theory}
The origin of String Theory lies in two observed features. The first is
the Regge trajectories, $J \propto M^2$, which define the higher spin
resonances, and which implies a finite size or spread in space for them.\\
The second feature is the dual resonance or $s-t$ channel scattering, which
lead to the Venetziano model. It was realized that if the
particles could be given an extension of the order of the Compton wavelength,
that is treated as strings, then the puzzling $s-t$ feature could be explained.\\
All this leads to strings which are governed by the equation\cite{r2}
\begin{equation}
\rho \ddot{y} - Ty'' = 0\label{e1}
\end{equation}
where the frequency $\omega$ is given by
\begin{equation}
\omega = \frac{\pi}{2} \sqrt{\frac{T}{\rho}}\label{e2}
\end{equation}
\begin{equation}
T = \frac{mc^2}{l}; \rho = \frac{m}{l}\label{e3}
\end{equation}
\begin{equation}
\sqrt{T/\rho} = c\label{e4}
\end{equation}
$T$ being the tension of the string, $l$ its length and $\rho$ the line density.
The identification (\ref{e3}) gives (\ref{e4}) where $c$ is the velocity of light,
and (\ref{e1}) then goes over to the usual d'Alembertian or massless Klein-Gordon equation. It is worth
noting that as $l \to 0$ the potential energy which is $\sim \int^l_0 T \left(\frac{\partial y}
{\partial x}\right)^2 dx$ rapidly oscillates.\\
Further if the above string is quantized canonically, we get
\begin{equation}
\langle \Delta x^2 \rangle \sim l^2\label{e5}
\end{equation}
The string effectively shows up as an infinite collection of Harmonic Oscillators
\cite{r2}.
It must be mentioned that (\ref{e5}) and (\ref{e3}) both show that $l$ is of the
order of the Compton wavelength.
This has been called one of the miracles of String theory by Veniziano\cite{r3}.
Infact the minimum length $l$ turns out to be given by $T/\hbar^2 = c/l^2$ which
from (\ref{e3}) and (\ref{e4}) is seen to give the Compton wavelength.\\
If the relativistic quantized string is given rotation\cite{r4}, then we
get back the equation for the Regge trajectories given above. Here we are dealing
with objects of finite extension rotating with the velocity of light rather like
spinning Black Holes. It must
be pointed out that in Super String theory, there is an additional term $a_0$
\begin{equation}
J \leq (2\pi T)^{-1}M^2 + a_0\hbar \quad \mbox{with}\quad a_0 = +1(+2)
\mbox{for the open (closed) string}\label{ex}
\end{equation}
In equation (\ref{ex}) $a_0$ comes from a zero point energy effect. When $a_0 = 1$
we have the usual guage bosons and when $a_0 = 2$ we have the gravitons.\\
It must be mentioned that in the relativistic case, even in Classical Theory it is known that\cite{r5,r6} there is
an extension of the order of the  Compton wavelength $l$ seen in
(\ref{e3}) above such that within this extension there are negative energies,
reminiscent of a Dirac particle with zitterbewegung, and indicative of the
breakdown of the concept of space time points.\\
Finally, if we go over to the full theory of Quantum Superstrings\cite{r7,r8}, we
deal with the Planck length, a non-commutative geometry and a total of ten
space time dimensions, six of which are curled up in the Kaluza-Klein sense.
\section{Quantized Fractal Space time}
A starting point for QFST is the observation that the electron can be
considered to be a Kerr-Newman Black Hole, which is otherwise classical, but
which also describes the purely QM anomalous $g=2$ factor\cite{r9}. However there is
a naked singularity, in other words the radius of the horizon becomes complex
and is given by
\begin{equation}
r_+ = \frac{GM}{c^2} + \imath b, b \equiv \left(\frac{G^2Q^2}{c^8} + a^2 -
\frac{G^2M^2}{c^4}\right)^{1/2}\label{e6}
\end{equation}
where $a$ is the angular momentum per unit mass.\\
The same circumstance is observed in the purely QM description of the electron
by the Dirac equation. This time we have
\begin{equation}
x = (c^2 p_1 H^{-1}t+a_1) + \frac{\imath}{2} c\hbar (\alpha_1 - cp_1H^{-1})H^{-1},\label{e7}
\end{equation}
The bridge between equations (\ref{e6}) and (\ref{e7}) is the fact that the imaginary
term in (\ref{e7}) is the zitterbewegung term which again is symptomatic of the fact
that space time points are not meaningful. Infact an average over Compton
scales removes the naked singularity.\\
Another way of looking at this is that for the Dirac or Klein-Gordan operator
we have a non Hermitian position operator given by
\begin{equation}
\vec X_{op} = \vec x_{op} - \frac{\imath \hbar c^2}{2} \frac{\vec p}{E^2}\label{ey}
\end{equation}
The Hermitianization of (\ref{ey}) leads to\cite{r10} a quantization of space
time at the Compton scale. As with strings, we have in this description a background
ZPF \cite{r10} corresponding to a collection of oscillation bounded by the
Compton wavelength. Interestingly an equation like (\ref{e2}) holds\cite{r6}.\\
Thus we have to deal with space time intervals, which are a sort of a minimum
cut off. This leads to the non commutative geometry
\begin{equation}
[x,y] = 0(l^2),[x,p_x] = \imath \hbar [1 + l^2], [t,E] = \imath \hbar [1+\tau^2]\label{e8}
\end{equation}
\cite{r11,r12}. Interestingly using (\ref{e8}) we can get back the Dirac representation.
Next, from the Dirac equation we can get the Klein-Gordon equation \cite{r13,r14}
or alternatively the massless string equation (\ref{e1}) subject to
(\ref{e4}). (Physically this means that bosons are bound states of Fermions.)
Further with bound Fermions we recover the equation of Regge trajectories
\cite{r15}. Interestingly, in Superstring theory also Dirac spinors
are introduced ad hoc, averages over an internal time $\tau$ are taken, and
then the Klein-Gordon equation of the bosonic theory is deduced\cite{r16}.
At the same time this space time fudge is exactly of the type required to explain the $s-t$
channel feature. It must be observed that in (\ref{e8}) if $l^2 = 0$,
then we are back with usual space time and Quantum Mechanics. It is only when
$l^2 \ne 0$ that we have a non commutativity, leading to Fermions. Indeed,
as also noted by Witten, our usual Lorentzian space time differentiable manifold
is bosonic and leads to the bosonic string or Klein-Gordon equation (\ref{e1}),
- but there is a substructure given by (\ref{e8}) which leads to the
Fermionic description and quantized space time. The former is an approximation
when $l^2 = 0$ and we have equation (\ref{e1}) and what Witten calls bosonic
space time (Cf. \cite{r17}). The latter gives the spinorial representation for the
Lorentz group which as noted by Einstein(Cf. ref.\cite{r18}) is the more
fundamental representation and also the Dirac equation, rather than (\ref{e1}).\\
It is quite significant that the above discrete space time features can be understood in terms of the
Nelsonian Theory in which the diffusion constant, in the light of the above
remarks has been meaningfully identified with $\hbar/m$ \cite{r19}. Here we are dealing
with a double Weiner process\cite{r20}, and the minimum space time cut offs
are stochastic in nature and may be called QFST.\\
Infact, a quick way to Quantum Mechanics from the Nelsonian theory is by starting with
the diffusion equation,
$$\Delta x \cdot \Delta x = \frac{h}{m} \Delta t$$
This can be rewritten as the usual Quantum Mechanical relation,
$$m \frac{\Delta x}{\Delta t} \cdot \Delta x = h = \Delta p \cdot \Delta x$$
To throw further light on this, we observe that it is well known\cite{r6} that from a two state formulation
we can recover the Schrodinger equation by considering the diffusion of a
particle to neighbouring points on either side, separated by a distance of the
order of the Compton wavelength. Similarly if we could consider a diffusion
forwards and backwards in time through a Compton time interval reminiscent of
the double Weiner process referred to, we can get back the d'Alembertian
equation, or the string equation (\ref{e1}). To see this more explicitly, let
us consider a point $x$ and its neighbouring points $x \pm l$ and a time
$t \pm \tau$. Then a typical simple diffusion equation would be given by
\cite{r21}, following Feynman,
\begin{equation}
C_\imath (t-\tau ) + C_\imath (t + \tau) = \sum_{j} [\delta_{\imath j} -
\frac{\imath}{\hbar}H_{\imath j} (t)\tau] C_j(t)\label{ez}
\end{equation}
where we denote by, $C_\imath(t) \equiv < \imath | \psi (t) >,$ the amplitude for
the state $| \psi (t) >$ to be in the state $| \imath >,$ and
$$< \imath | U|j > \equiv U_{\imath j} (t + \tau t, t) \equiv \delta_{\imath j}
- \frac{\imath}{\hbar}H_{\imath j}(t)\tau t.$$
and $\imath$ denotes the point $x$ and the points $x \pm l$. In the limit
$l = c \tau \to 0$ (\ref{ez}) gives the Klein-Gordon equation or alternatively the
equation (\ref{e1}) subject to the condition (\ref{e4}). This also shows the close
relation of the diffusion process and special relativity, as noted in \cite{r19}.\\
Finally in the above formulation using the spirit of fluctuations and the non commutative relation (\ref{e8}),
we get, $\Delta x \sim l^3$, where $l$ is the electron Compton wavelength\cite{r20}.
We are thus lead firstly to the Planck length and secondly to the fact that
the single $x$ dimension has now become three dimensional. The two extra
dimensions are fluctuations and curled up in the Kaluza-Klein sense\cite{r22} as shown elsewhere\cite{r20}.
It must be stressed that in this formulation we use as seen in the Super String
case, the zero point fluctuations.\\
\section{Conclusion}
It is not surprising that QSS and QFST reach the same extra
dimensions, non commutative geometry and Planck scale, because the former
starts with extension and then the QM input, while the latter starts with
extension and the Nelsonian QM input. From this point of view we get an
insight into the otherwise inexplicable features of QSS. As Veneziano put it
\cite{r3}, "we thus face a kind of paradoxical situation. On the other hand
quantum mechanics is essential to the success of the KK idea. At the same time,
QFT gives meaningless infinities and spoils the nice semiclassical results. If
the beautiful KK idea is to be saved we need a better quantum theory than QFT.
I will argue below that such a theory already exists: it is called
(super) string theory." The key lies in
space time intervals which underlie a Fermionic sub structure of space time given
by (\ref{e8}), which has a perfectly natural origin in stochastic process or
in the spirit of which lies "Law without Law"\cite{r23}, and is exemplified by
QFST.


\begin{thebibliography}{99}
\bibitem {r1} F.D. Peat, "Superstrings", Abacus, Chicago, 1988, p.21.
\bibitem {r2} G. Fogleman, Am.J.Phys., 55(4), 1987, pp.330-336.
\bibitem {r3} G. Veneziano, "Quantum Geometric Origin of All Forces in
String Theory" in "The Geometric Universe", Eds. S.A. Huggett et al., Oxford
University Press, Oxford, 1998, pp.235-243.
\bibitem {r4} J. Scherk, Rev.Mod.Phys., 47 (1), 1975, pp.1-3ff.
\bibitem {r5} C. Moller, "The Theory of Relativity", Clarendon Press,
Oxford, 1952, pp.170 ff.
\bibitem {r6} B.G. Sidharth, Ind.J. Pure \& Appld.Phys., Vol.35, 1997, pp.456-471.
\bibitem {r7} W. Witten, Physics Today, April 1996, pp.24-30.
\bibitem {r8} Y. Ne'eman, in Proceedings of the First Internatioinal Symposium,
"Frontiers of Fundmental Physics", Eds. B.G. Sidharth and A. Burinskii,
Universities Press, Hyderabad, 1999, pp.83ff.
\bibitem {r9} C.W. Misner, K.S. Thorne and J.A. Wheeler, "Gravitation",
W.H. Freeman, San Francisco, 1973, pp.819ff.
\bibitem {r10} B.G. Sidharth, Int.J.Mod.Phys.A, 13 (15), 1998, p.2599ff.
\bibitem {r11} H.S. Snyder, Physical Review, Vol.72, No.1, July 1 1947, p.68-71.
\bibitem {r12} B.G. Sidharth, Chaos, Solitons and Fractals, 11 (2000), 1269-1278.
\bibitem {r13} V. Heine, "Group Theory in Quantum Mechanics", Pergamon Press,
Oxford, 1960, p.364.
\bibitem {r14} J.R.Klauder, "Bosons Without Bosons" in Quantum Theory
and The Structures of Time and Space, Vol.3 Eds by L. Castell, C.F. Van
Weiizsecker, Carl Hanser Verlag, Munchen 1979.
\bibitem {r15} B.G. Sidharth, "Scaled Universe II", to appear in Chaos, Solitons and
Fractals.
\bibitem {r16} P. Ramond, Phys.Rev.D., 3(10), 1971, pp.2415-2418.
\bibitem {r17} B.G. Sidharth "A Brief Note on Analaticity and Causality, and
thee 'Levels of Physics'", to appear in Chaos, Solitons and Fractals.
\bibitem {r18} M. Sachs, "General Relativity and Matter", D. Reidel
Publishing Company, Holland, 1982, p.45ff.
\bibitem {r19} B.G. Sidharth, Chaos, Solitons and Fractals, 12(1), 2000, 173-178.
\bibitem {r20} B.G. Sidharth, "Unification of Electromagnetism in Quantized
Fractal Space Time", to appear in Chaos, Solitons and Fractals.
\bibitem{r21} R.P. Feynman, The Feynman Lectures on Physics, 2, Addison-Wesley,
Mass., 1965.
\bibitem {r22} H.C. Lee, "An Introduction to Kaluza-Klein Theories", World
Scientific, Singapore, 1984.
\bibitem {r23} J.A. Wheeler, Am.J.Phys., 51(5), 1983, p.398.
\end{thebibliography}
\end{document}